# Photoresponse of a strongly correlated material determined by scanning photocurrent microscopy

T. Serkan Kasırga[1], Dong Sun[1], Jae H. Park[1], Jim M. Coy[1], Zaiyao Fei[1], Xiaodong Xu[1,2*], David H. Cobden[1*]

*[1]Department of Physics, University of Washington, Seattle WA 98195, USA*
*[2]Department of Materials Science and Engineering, University of Washington, Seattle WA 98195, USA*

*Corresponding authors: xuxd@uw.edu, cobden@uw.edu

**The generation of a current by light is a key process in optoelectronic and photovoltaic devices. In band semiconductors, depletion fields associated with interfaces separate long-lived photo-induced carriers. However, in systems with strong electron-electron and electron-phonon correlations it is unclear what physics will dominate the photoresponse. Here we investigate photocurrent in a vanadium dioxide, an exemplary strongly correlated material known for its dramatic metal-insulator transition[1-3] (MIT) at $T_c \approx 68$ °C which could be useful for optoelectronic detection and switching up to ultraviolet wavelengths[4-10]. Using scanning photocurrent microscopy (SPCM) on individual suspended VO₂ nanobeams we observe photoresponse peaked at the metal-insulator boundary but extended throughout both insulating and metallic phases. We determine that the response is photo-thermal, implying efficient carrier relaxation to a local equilibrium in a manner consistent with strong correlations[11-14]. Temperature dependent measurements reveal subtle phase changes within the insulating state. We further demonstrate switching of the photocurrent by optical control of the metal-insulator boundary arrangement. Our work shows the value of SPCM applied to nanoscale crystals for investigating strongly correlated materials, and the results are relevant for designing and controlling optoelectronic devices employing such materials.**

Strongly correlated materials offer the potential for applications beyond the limits of semiconductor technologies, but their complex nature makes it challenging to determine the fundamental mechanisms behind their behavior. In the case of VO₂, recent progress[15-27] in working with crystals smaller than the characteristic domain size has allowed clarification of a number of aspects which were obscured by domain structure and other crystal imperfections in earlier bulk studies. These include improved measurements of the resistivity[18-20], the activation energy and the optical gap in the insulator[26]; the existence of a threshold resistivity for the transition[18,20]; and improved understanding of the interplay between the two similar monoclinic insulating (I) phases, M1 and M2 (their structures are indicated in Fig. 1), alongside the rutile metallic (M) phase near the MIT[20-25]. The resulting improved level of control and understanding of VO₂ now presents the opportunity to investigate methodically the optoelectronic response of a strongly correlated electronic material.

We applied SPCM (see Methods) at a wavelength of 800 nm (1.55 eV, well above the 0.6 eV optical gap in the insulator) to suspended VO₂ nanobeam devices, as depicted in Fig. 1. Suspending the nanobeams removes complications from nonuniform stress caused by substrate adhesion[17,18], but an axial stress is still present due to firm attachment under the contacts. Fig. 2 shows measurements at low laser power $P = 1.0$ μW (~20 W/cm²) for two similar nanobeams. In the top row (Figs. 2a-c) are grayscale plots of reflected intensity vs. laser position. Device 1, at 30 °C, is well below $T_c$ and is uniformly in the M1 insulating ($I_{M1}$) phase, as determined by Raman spectroscopy and resistivity measurements (see Supplementary Materials). Device 2, at 75 °C and 90 °C, is above $T_c$ and shows a darker metallic (M) region in coexistence with a paler insulating region. The latter is in the M2 phase ($I_{M2}$), because the axial tension stabilizes the insulating phase with the longer c-axis lattice constant[28] (see Fig. 1). We orient the devices with the insulating region on the left, the bias $V$ being applied to the left contact and the photocurrent $I_{ph}$ measured out of the right contact. In the second row are corresponding colorscale maps of the zero-bias photocurrent, $I_0$. Below $T_c$ (Fig. 2d) $I_0$ is very small, while above $T_c$ (Figs. 2e,f) it is much larger, positive, and extends along the entire nanobeam reaching a maximum at the I-M boundary. Figs. 2g-i show the variation of $I_0$ along the center-line of the nanobeam (thick black line), along with the variation of $I_{ph}$ measured at several finite biases. The change in current due to the bias can be described entirely in terms of a bias-independent photoconductance $G_{ph} = (I_{ph} - I_0)/V$, as shown in Figs. 2j-l.

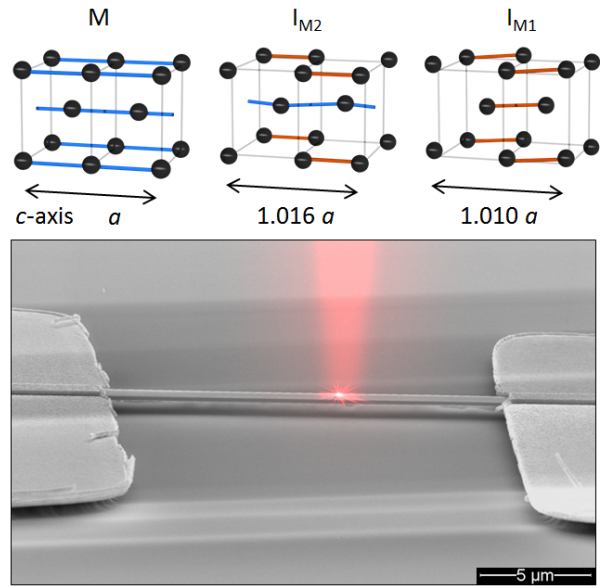

**Figure 1 | Scanning photocurrent microscopy on a suspended VO₂ nanobeam device.** A focused laser is depicted superimposed on an SEM image of a VO₂ nanobeam suspended between gold electrical contacts. Above are sketches of the arrangement of V atoms in the tetragonal metallic phase (M) and monoclinic insulating phases ($I_{M1}$ and $I_{M2}$). Gray lines indicate two tetragonal unit cells; O atoms are not shown. The M phase is shortest along the tetragonal c-axis, which is the nanobeam axis, and has only periodic V chains (blue lines); $I_{M2}$ is longest and has half the V atoms dimerized (red lines); and $I_{M1}$ has intermediate length and all V atoms dimerized.

To understand these measurements, we begin by exploiting the phase transition itself to quantify the rise in temperature under the laser spot by observing the position of the I-M boundary[18] above $T_c$. The fraction $x_b$ of insulator ($I_{M2}$) in the nanobeam depends on the lattice temperature $T_b$ at the boundary, because $x_b$ determines the axial tension which must be appropriate for the two phases to coexist at $T_b$. $x_b$ decreases with increasing stage temperature $T_0$, as indicated by the red circles in Fig. 3a. It also decreases with increasing laser power $P$ at fixed $T_0$, as shown by the black circles. By comparing the effects of increasing $P$ and $T_0$, as illustrated in Fig. 3b, we deduce that for Device 1 the local temperature rise of the lattice with the laser near the middle is ≈1.5 °C per μW of laser power.

At higher $P$ the interface has a curved appearance (see Fig. 3b, bottom image). This is because the power absorbed is greatest when the laser is focused on the center-line of the nanobeam, and hence the temperature rise and the boundary shift is also greater. In fact, as $P$ increases the arrangement of the phases is increasingly disturbed by the laser, and when $P > \sim 10$ μW the metallic region is dragged along by the laser beam leading to complex nonlinear behavior. We therefore confine our studies to lower power levels where the response is linear in $P$.

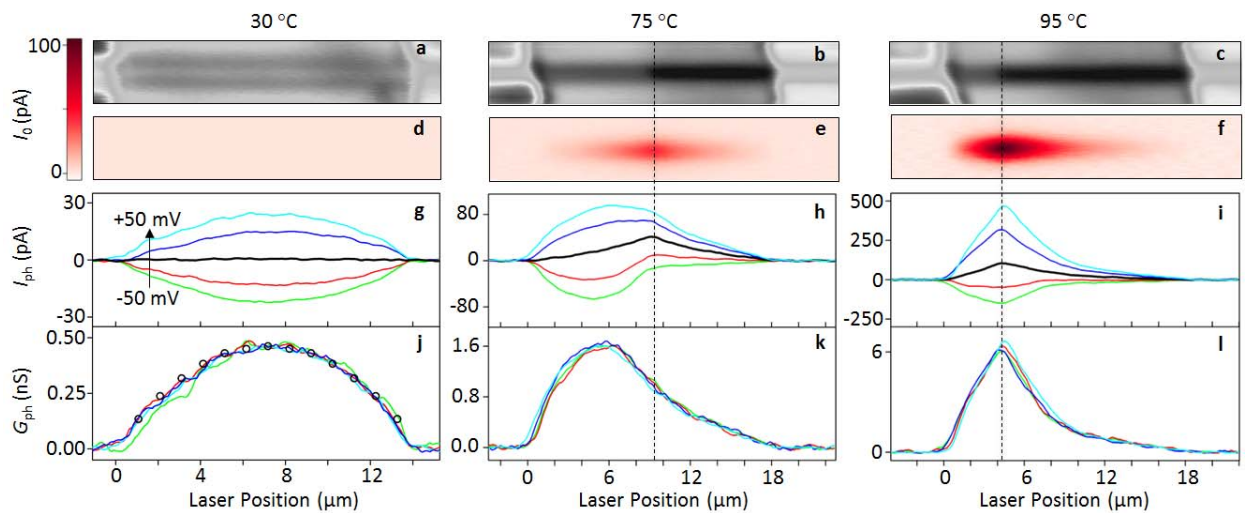

**Figure 2 | Photocurrent measurements at temperatures across the metal-insulator transition.** Data at 30 °C (left column) are from one nanobeam (Device 1); those at 75 and 95 °C (center and right columns) are from another very similar nanobeam (Device 2). The laser power was $P = 1.0$ μW. **a-c,** Reflected light images. The metallic phase appears darker. **d-f,** Corresponding zero-bias photocurrent images. **g-i,** Photocurrent $I_{ph}$ along the center-line of the nanobeam for voltage biases $V$ = -50,- 30, 0 ($I_0$, in black), 30 and 50 mV as indicated. **j-l,** Photoconductance $G_{ph} = (I_{ph} - I_0)/V$, which can be seen to independent of $V$. Circles in **j** plot the function $G_0 x_l (1 - x_l)$ predicted for a thermal resistance change (see text), where $x_l$ is fractional position along the nanobeam and fitting parameter $G_0 = 1.9$ nS.

Using this knowledge of the temperature rise we consider next the photoconductance. Well below $T_c$, when the nanobeam is entirely in the I$_{M1}$ phase (Fig. 2j), the negative temperature coefficient of the insulator resistivity $\rho_I(T)$ will lead to a conductance increase $\triangle G$ under illumination. For a small temperature rise $\delta T(x)$, where coordinate $x$ runs from 0 at the left contact to 1 at the right, $\triangle G \approx -\frac{1}{R^2} \frac{dR}{dT} \delta T_{av}$, where $\delta T_{av} = \int_0^1 \delta T(x) \, dx$, and $R(T)$ is the resistance of the nanobeam held at uniform temperature $T$ (see Supplementary Materials for details). If we assume that all the heat flows along the nanobeam, then $\delta T(x)$ drops linearly to zero at the gold contacts and $\delta T_{av} = \delta T(x_l)/2 \propto x_l(1 - x_l)$, where $x_l$ is the laser position. For the laser at 1.0 μW in the middle we know from above that $\delta T(x_l = 1/2) \approx 1.5$ °C, and using the independently measured dark resistance $R = 42$ MΩ and $dR/dT = 1.4$ MΩ/°C we obtain $\triangle G \approx 0.6$ nS. The measured photoconductance $G_{ph}$= 0.5 nS at $x_l = 1/2$ is slightly smaller than this, which is explained by heat loss through the air making $\delta T_{av}$ smaller than $\delta T(x_l)/2$ (see below). In addition, the predicted variation with laser position, $\triangle G \propto x_l(1 - x_l)$, is an excellent match to the experimentally determined $G_{ph}$, as shown by the open circles in Fig. 2j.

Above $T_c$ when an I-M boundary is present (Figs. 2k and l), the decrease in the insulating fraction $x_b$ with boundary temperature rise $\delta T_b \equiv T_b - T_0$ results in a conductance increase which is largest when $\delta T_b$ is maximum, ie, when $x_l = x_b$ (see Supplementary Materials). This explains the fact that $G_{ph}$ is peaked at the boundary at 95 °C. More complex behavior results when the effect of the change in $\rho_I(T)$ is comparable, as is the case at 75 °C. In summary, both above and below $T_c$ the measured photoconductance can be well understood as the result of the temperature rise of the lattice with no hint of any other contribution.

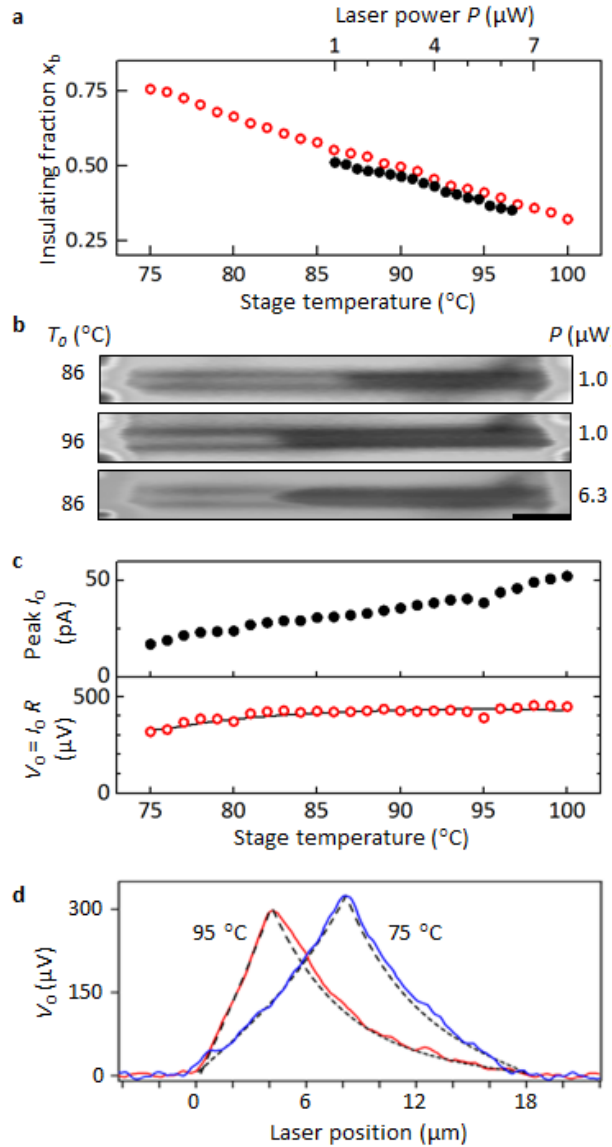

**Figure 3 | Evidence for the thermoelectric origin of the photocurrent. a,** Comparison of the variation of insulating fraction $x_b$ with stage temperature $T_0$ at laser power $P = 1.0$ μW (red open circles), and with $P$ at $T_0 = 86$ °C (black filled circles), for Device 1. **b,** Reflection images (generated by the scanning laser spot) illustrating the comparison. The scale bar is 2 μm. **c,** Temperature dependence of the peak zero-bias photocurrent $I_0$ (upper panel, black filled circles) and emf $V_0 = I_0 R$ (lower panel, red open circles) for Device 1. **d,** Variation of $V_0$ with laser position for Device 2 at two temperatures. The solid line in **c** and the dashed lines in **d** are the predicted photothermal emf (see text).

We are now in a position to analyze the zero-bias photocurrent $I_0$ seen above $T_c$ (Figs. 2e,f). If carrier relaxation to complete equilibrium with the lattice is fast then $I_0$ will be purely photothermal (thermoelectric), so we consider this possibility first. The lattice temperature difference ($\delta T_b$) between the I-M boundary and the gold contacts will generate a thermoelectric emf, $V_{te} = -\Delta S_{IM} \delta T_b$, due to the difference in Seebeck coefficients, $\Delta S_{IM} = S_I - S_M$, between the $I_{M2}$ and M phases. From the literature[27,29] $S_I \approx$ -350 µV/°C (for $I_{M2}$) and $S_M \approx$ -20 µV/°C, so $\Delta S_{IM} \approx$ -330 µV/°C. Hence when the laser at 1 µW is focused near the middle, giving $\delta T_b \approx$ 1.5 °C for Device 1 as found above, we expect $V_{te} \approx$ +500 µV. Fig. 3c shows the temperature dependence of the peak value of $I_0$ and the corresponding emf, $V_0 = I_0 R$. $V_0$ reaches $\approx$ +450 µV, in excellent agreement with $V_{te}$ considering the uncertainty in knowledge of the thermoelectric coefficients.

Moreover, we can calculate the variation of $V_{te}$ with laser position, allowing for different thermal conductivities $\kappa_M$ and $\kappa_{I_{M2}}$ of the two phases and for heat loss $\beta \delta T$ through the air (see Supplementary Information). Fig. 3d shows the measured variation of $V_0$ with laser position (solid lines) for Device 2 at 75 and 95 °C compared with the results of the calculation of $V_{te}$ (dashed lines). Here we used[30] $\kappa_I = 3.5$ W/m/°C and $\kappa_M/\kappa_I = 2$ and treated $\beta$ and the fraction $\gamma$ of the laser power absorbed as fitting parameters, yielding $\beta = 0.03$ W/°C/m, consistent with the thermal conductivity of air, and $\gamma = 0.5$. We also calculated the temperature variation of the peak value of $V_{te}$, which occurs for the laser at the I-M boundary ($x_l = x_b$), again obtaining excellent agreement with the temperature dependence of $V_0$ (see the dashed line in Fig. 3c; for this device $\gamma = 0.6$). We conclude that the dominant photocurrent contribution follows directly from the lattice temperature rise just as does the photoconductance.

Having established that the photocurrent is predominantly photothermal, we now address the question of whether there is any additional contribution from separation of nonequilibrium carriers which diffuse to the I-M interface, as might be expected in a band semiconductor like silicon[31]. Such a contribution is not possible for excitation in the metallic phase because electron-lattice relaxation occurs in picoseconds in all metals (there being no gap to block low-energy processes). Thus the photocurrent signal seen extending many microns into the metallic region (Fig. 2h) must be entirely photothermal. Since the photothermal mechanism consistently explains the entire observed photocurrent in both metallic and insulating parts equally well, we deduce that any additional contribution in the insulator is insignificant. This is consistent with efficient local carrier relaxation in insulating $VO_2$ which keeps the material very close to local thermal equilibrium during illumination. Evidence for fast relaxation is provided by the fact that the photocurrent we measure is identical for pulsed (0.25 ps pulses repeated at 76 MHz) and for continuous wave excitation at the same average power. In addition, we saw exactly the same behavior using different laser wavelengths, consistent with only the absorbed power being relevant (see Supplementary Materials). We note that there are no reports of nonequilibrium carrier effects, such as luminescence, in the optical response of insulating $VO_2$ in the literature. Moreover, the strong electron-phonon coupling and polaronic effects which are likely in such a material provide a natural mechanism for efficient relaxation, through interband scattering and slow diffusion. This is congruent with the very short scattering length, roughly equal to the lattice constant, that can be inferred from the poor conductivity of the metallic phase and the prefactor of the activated insulator conductivity.

The photothermal origin of the photocurrent is further supported by a number of other observations. One is that if the metallic region is pulled into the center of the nanobeam using a second identical fixed laser spot, resulting in two opposing I-M boundaries, then $I_0$ almost vanishes for all positions of the scanning laser (Fig. 4a). This follows from the fact that the temperatures at both boundaries must be the same for the two phases to coexist under the same axial strain: hence the boundaries generate emfs of equal magnitude but opposite sign and the sum vanishes. Such a vanishing of the photoresponse, independent of laser position, would not occur for other mechanisms. Another example is that for a nanobeam not released from the substrate by etching, multiple alternating I and M domains occur due to inhomogeneous strain[17], and we see associated peaks in $I_0$ of alternating sign centered at the I-M boundaries (Fig. 4b). This is explained by the fact that in this case thermal conduction through the substrate causes the temperature rise to be much more localized to the laser spot so that the thermoelectric emf is large only when the laser is close to an I-M boundary.

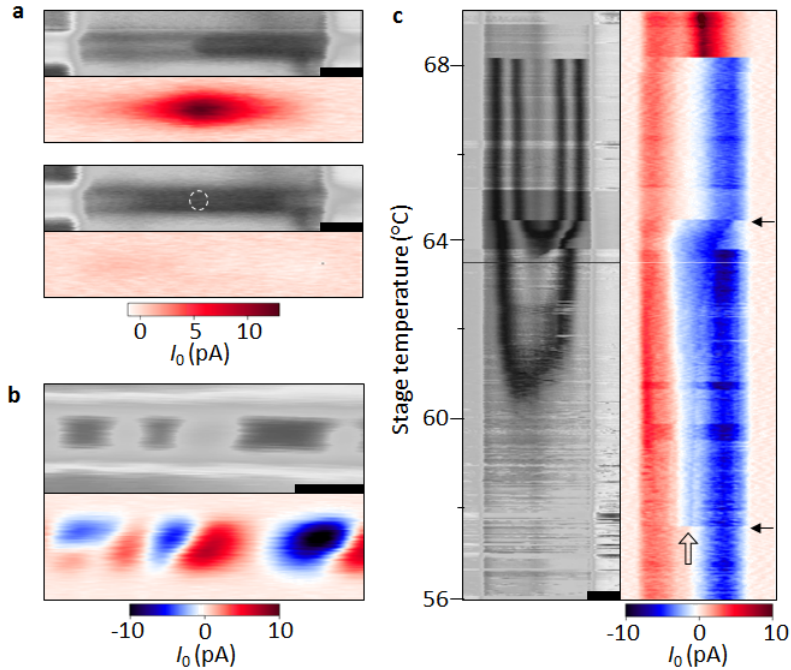

**Figure 4 | Optical control of the photocurrent and imaging the evolution of the insulating phase. a,** Reflection and zero-bias photocurrent images of Device 1 at 70 °C using $P = 1.0$ μW, without (above) and with (below) a second laser ($P = 2.4$ μW) focused at the location of the dashed circle. **b,** Similar images of an unsuspended nanobeam ($P = 0.6$ μW). **c,** Reflection (left) and photocurrent (right) measured by repeatedly scanning along the center-line of a nanobeam (Device 3) while ramping the temperature up to and beyond the transition at 68 °C. Horizontal arrows indicate where the conversion of $I_{M1}$ to $I_{M2}$ begins (lower) and completes (upper). Scale bars are 2 μm in **a** and **b**, 5 μm in **c**.

Finally, we illustrate how SPCM can be exploited to study such a solid-state phase transition by allowing the visualization of interphase boundaries that are otherwise invisible. For example, the interconversion between $I_{M1}$ and $I_{M2}$ near the MIT[21-25] may be revealed even though the two phases are optically almost indistinguishable. Fig. 4c shows the reflection signal (left) and zero-bias photocurrent (right) measured while repeatedly scanning along the center-line of a nanobeam as the temperature is ramped up. Above 68.2 °C the nanobeam is straight and in $I_{M2}$-M coexistence, and $I_0$ peaks at the boundary. Below 57.5 °C it is straight and fully $I_{M1}$. As the temperature rises from 57.5 to 64.5 °C (between the two horizontal arrows) a stripe pattern develops in the reflection signal because the nanobeam gradually becomes buckled due to conversion from $I_{M1}$ to $I_{M2}$ with its longer c-axis. During this process a feature appears in the photocurrent near the middle of the nanobeam (white arrow) and steadily expands. The feature may reflect an $I_{M2}$ domain nucleating and growing, made visible by the difference in photoresponse between $I_{M2}$ and $I_{M1}$.

In summary, we have determined the relationship between the optical and dc electrical properties of $VO_2$ using scanning photocurrent microscopy, which probes the optoelectronic properties of the phases and their interfaces. We observe photoconductance and zero-bias photocurrent generation which is entirely of photothermal origin, consistent with very efficient electron-lattice relaxation in the strongly correlated insulating phase and in stark contrast with the response of uncorrelated band insulators.

**Methods**

To perform SPCM, a diffraction-limited 800 nm laser spot, chopped at 1 kHz, is scanned over the sample on a heated stage in air, and the resulting photocurrent is measured with a current preamplifier and a lock-in amplifier referenced to the chopper. The reflected laser light is detected by a silicon photodiode, generating an image of the device corresponding directly to the photocurrent image. The nanobeams are grown by physical vapor transport[16] using a $V_2O_5$ source placed in an alumina crucible in the center of a tube furnace at ~900 °C and argon carrier gas at a few mbar. The substrate is a p-doped (100) Si chip with a 2-μm wet oxide coating. The nanobeams grow elongated along the tetragonal c-axis with {110} sides[16]. The contacts (10 nm Ti followed by 200 nm Au) are made by photolithography followed by electron-beam evaporation and lift-off, and the nanobeams are suspended by immersing the devices in buffered oxide etch for several minutes.

**Acknowledgments**

This work was supported by the U.S. Department of Energy, Office of Basic Energy Sciences, Division of Materials Sciences and Engineering, Award DE-SC0002197; by the Army Research Office, Contract 48385-PH; and by Xu's NSF Career Award DMR-1150719. We thank Boris Spivak for helpful discussions.

# Photoresponse of a strongly correlated material determined by scanning photocurrent microscopy

T. Serkan Kasırga[1], Dong Sun[1], Jae H. Park[1], Jim M. Coy[1], Zaiyao Fei[1], Xiaodong Xu[1,2*], David H. Cobden[1*]

[1]*Department of Physics, University of Washington, Seattle WA 98195, USA*

[2]*Department of Materials Science and Engineering, University of Washington, Seattle WA 98195, USA*

[*]Corresponding author emails: cobden@uw.edu and xuxd@uw.edu

**Table of Contents**

1. Identity of the phases
2. Wavelength dependence
3. Photoconductance calculations
4. Calculation of the metal-insulator boundary temperature rise
5. Calculation of the thermoelectric peak emf vs stage temperature

## Supplementary Text

### 1. Identification of the phases

The identity of the M (rutile), $I_{M1}$ and $I_{M2}$ phases can be confirmed by their resistivity and Raman spectra. All the suspended nanobeams in this study were fully $I_{M1}$ at room temperature while above $T_c$ ~68 °C they were in coexistence between the $I_{M2}$ and M phases. A representative resistance-temperature measurement (in the dark) is shown in Fig. S1a. The resistance actually *increases* as the device is warmed through the transition. This is because $I_{M1}$ converts to $I_{M2}$ due to the axial tension above $T_c$ and $I_{M2}$ has approximately three times higher resistivity than $I_{M1}$. As we showed in an earlier paper[1], in coexistence the paler looking $I_{M2}$ part of the nanobeam has resistivity 12 $\Omega$cm and dominates the total resistance; the resistance of the metallic domain is negligible. At higher temperatures, up to abuot 105 °C, the resistance steadily decreases as the $I_{M2}$ domain shrinks. The identity of each insulating phase was confirmed by Raman spectroscopy *in situ*: Fig S1b shows examples of Raman spectra. The peak above 600 cm$^{-1}$ allows straightforward distinction between $I_{M1}$ and $I_{M2}$ (see eg ref.2).

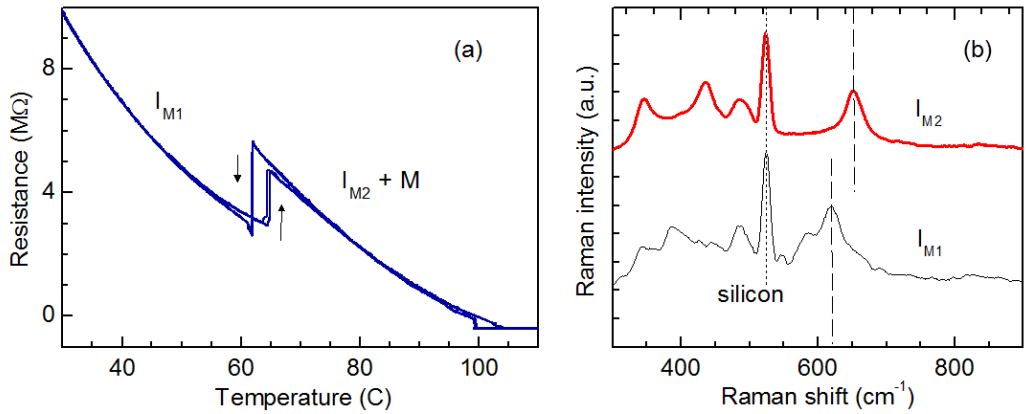

**Figure S1 | a**, Resistance vs temperature for a suspended VO$_2$ nanobeam over two temperature cycles. b, Raman spectra obtained from a nanobeam at room temperature (in the I$_{M1}$ phase, black) and the insulating part of a nanobeam above $T_c$ (in the I$_{M2}$ phase, red). The different position of the characteristic peak a little above 600 cm$^{-1}$ in I$_{M1}$ and I$_{M2}$ is indicated by dashed lines. The silicon substrate also produces a strong peak (dotted line).

## 2. Dependence on laser wavelength

The measured behavior was found to be very similar for different laser wavelengths, consistent with a photothermal mechanism in which only the absorbed laser power is relevant. In particular, we did a number of measurements using a green laser. For example, in Fig. S2 we compare line traces of the zero-bias photocurrent $I_0$ measured on the same device in coexistence at wavelengths of 532 nm and 800 nm. Different powers were used for the two lasers. The photocurrent profile along the nanobeam is essentially identical when scaled by the power difference (implying that the absorption coefficient at both wavelengths is similar.) The bump in the insulating phase (laser position ~4 μm) is related to the relatively high laser power here, which is large enough to perturb the domain arrangement somewhat.

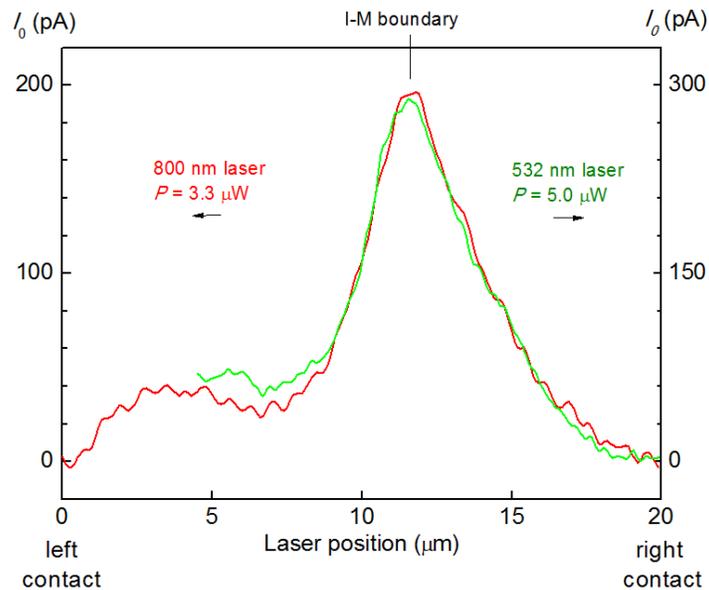

**Figure S2 |** Zero-bias photocurrent vs laser position measured along the center-line of a nanobeam (Device 1) in coexistence at 70 °C for two different laser wavelengths and powers.

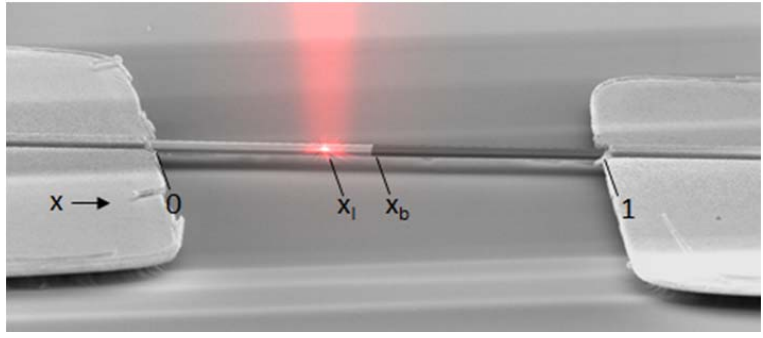

**Figure S3** | SEM image with definitions (for the case $x_l < x_b$). The nanobeam has been colored dark on the right to indicate a metallic domain, though in reality there is no I-M contrast in the electron microscope.

## 3. Photoconductance calculations

We define $x$ to be the fractional position along the suspended part of the nanobeam, running from 0 at the left contact to 1 at the right contact. (This corresponds to a length between 15 and 20 µm, depending on device). In I-M coexistence we define the insulating region to be on the left. The position of the I-M boundary, $x_b$, is then equal to the fraction of the nanobeam in the insulating phase. The laser is focused at position $x_l$, as indicated in Figure S3. The temperature at point $x$ is $T(x) = T_0 + \delta T(x)$, where $T_0$ is the stage temperature. The measured change in conductance when the laser is applied is $\Delta G = \Delta\left(\frac{1}{R}\right) = \frac{1}{R(laser\ on)} - \frac{1}{R(laser\ off)}$, where $R$ is the resistance of the nanobeam. We consider here what happens if $\Delta G$ is determined solely by the temperature rise $\delta T(x)$.

For a nanobeam entirely in one insulating phase, and for small $\delta T$,

$$\Delta G = \Delta\left(\frac{1}{R}\right) \approx -\frac{1}{R^2}\Delta R \approx -\frac{1}{R^2}\frac{L}{A}\int_0^1 \frac{d\rho}{dT}\delta T(x)\,dx \approx -\frac{1}{R^2}\frac{L}{A}\frac{d\rho}{dT}\int_0^1 \delta T(x)\,dx = -\frac{1}{R^2}\frac{dR}{dT}\delta T_{av}\,,$$

as stated in the main text. Here $A$ is the cross-sectional area of the nanobeam (assumed uniform), $L$ is the distance between the contacts, $\rho(T)$ is the resistivity of the insulating phase, and $R(T) = \frac{\rho(T)L}{A}$ is the resistance when the nanobeam is held at uniform temperature $T$. Thus the negative temperature coefficient of $\rho(T)$ leads to a positive $\Delta G$ proportional to the average temperature rise $\delta T_{av} = \int_0^1 \delta T(x)\,dx$ along the nanobeam, with a dependence on laser position that follows this quantity, $\Delta G(x_l) \propto \delta T_{av}(x_l)$.

To calculate $\delta T_{av}(x_l)$ we first assume that all the heat flows along the nanobeam. Then $\delta T(x)$ simply drops linearly from a maximum value of $\delta T_l = \delta T(x_l)$ at the laser spot to zero at each of the gold contacts and as a result $\delta T_{av} = \delta T_l/2$. By symmetry $\delta T_l$ is maximum when $x_l = 1/2$, and making the heat current to the left and right sum to a constant (the absorbed laser power) gives $\delta T_l \propto x_l(1 - x_l)$. Thus

$$\Delta G \propto \delta T_{av} \propto \delta T_l \propto x_l(1 - x_l)\,. \qquad (S1)$$

This variation fits the measurements of $G_{ph}$ vs $x_l$ very well, as seen in Fig. 2j. If some heat is lost by conduction through the air or radiation, as considered further in the next section, then the temperature rise is more localised to the laser spot and $\delta T_{av} < \delta T(x_l)/2$. The form of $\Delta G(x_l)$ also differs slightly from Eq. (S1), but the difference is indistinguishable in our measurements.

When an I-M boundary is present, $R$ is dominated by the insulating fraction of the nanobeam, as discussed in Section 1. There is then another contribution to $\Delta R$, because in addition to any decrease in the insulator resistivity there is also a decrease in the amount $x_b$ of insulator. $x_b$ changes linearly with the boundary temperature (since it is determined by the requirement that the axial stress be appropriate for the two phases to coexist at the I-M boundary temperature[1] $T_b$). This contribution is thus proportional to $\delta T_b = \delta T(x_b) = T_b - T_0$. It is largest when $\delta T_b$ is maximum, that is, when $x_l = x_b$, and it peaks when the laser is at the boundary, whereas the contribution from changing $\rho(T)$ peaks when the laser is focused on the insulating part. The behavior of $G_{ph}$ seen in Figs. 2k and l can be qualitatively explained this way, but is difficult to model accurately.

## 4. Calculation of the metal-insulator boundary temperature rise

The thermoelectric emf is $V_{te} = -\Delta S_{IM} \delta T_b$ so we need to find how $\delta T_b$ varies with $x_l$ in coexistence. For this we need to determine the temperature increase $\delta T(x)$ for given $x_l$, $x_b$, and laser power $P$. We assume $P$ is small enough that changes in $x_b$ can be ignored (they will give effects quadratic in $P$).

As before we assume that due to absorption of power $\gamma P$ ($\gamma$ is a constant) the temperature rises at the laser spot by $\delta T_l \equiv \delta T(x_l)$ and is unchanged at the gold contacts, $\delta T(0) = \delta T(1) = 0$. $\delta T$ will satisfy a one-dimensional steady-state heat equation of the form

$$\frac{\kappa A}{L^2} \frac{d^2 \delta T(x)}{dx^2} - \beta \delta T(x) = 0,$$

that is,

$$\frac{d^2 \delta T}{dx^2} - \alpha^2 \delta T = 0 \qquad (S2)$$

where $\alpha = \sqrt{\beta L^2 / \kappa A}$ and $\kappa$ is the c-axis thermal conductivity. Here $\beta \delta T$ is the rate of heat loss per unit length by conduction through the air to the substrate (and radiation, but this is negligible). The thermal conductivity is allowed to be different for the I and M phases: $\kappa = \kappa_I$ for $x < x_b$ and $\kappa = \kappa_M$ for $x > x_b$. One can also allow $\gamma$ to be different for I and M, but this creates a step in the response as the laser crosses the interface at $x_l = x_b$ which we do not see, so we take $\beta$ to have a single value.

For the case $x_l < x_b$ Eq. (S2) must be solved piecewise in the three regions $I$ $(0, x_l)$, $II$ $(x_l, x_b)$ and $III$ $(x_b, 1)$. By taking the general solution $\delta T = A e^{\alpha x} + B e^{-\alpha x}$, with the appropriate choice of $\alpha_I = \sqrt{\beta L^2 / \kappa_I A}$ or $\alpha_M = \sqrt{\beta L^2 / \kappa_M A}$ in each region, and matching the specified temperatures at the end of each region, we get

$$\delta T(x) = \begin{cases} \dfrac{\delta T_l \sinh \alpha_I x}{\sinh \alpha_I x_l} & 0 \leq x \leq x_l \\[3mm] \dfrac{1}{\sinh \alpha_I (x_l - x_b)} [\delta T_l \, \sinh \alpha_I (x - x_b) - \delta T_b \sinh \alpha_I (x - x_l)] & x_l \leq x \leq x_b \\[3mm] \dfrac{\delta T_b \sinh \alpha_M (1 - x)}{\sinh \alpha_M (1 - x_b)}. & x_b \leq x \leq 1 \end{cases}$$

The values of $\delta T_l$ and $\delta T_b$ can determined by applying the boundary conditions between regions $I$, $II$ and $III$ corresponding to conserving energy flow: at $x = x_l$ we have

$$\frac{\kappa_I A}{L}\left(\frac{dT}{dx}\Big|_{x_l,I} - \frac{dT}{dx}\Big|_{x_l,II}\right) = \gamma P \; ,$$

and at $x = x_b$ we have

$$\kappa_I \frac{dT}{dx}\Big|_{x_b,II} - \kappa_M \frac{dT}{dx}\Big|_{x_b,III} = 0 \; .$$

After making use of the expressions for $\delta T(x)$ above and eliminating $\delta T_l$ we obtain the required expression:

$$\delta T_b(x_l) = \frac{\delta T_l(x_l)}{C_1} = T_P\,\frac{\sinh\alpha_I(x_b - x_l)}{C_1 C_2 - 1}\;, \qquad (S3)$$

Where $T_P = \gamma P/\sqrt{\kappa_I \beta A}$ ,

$$C_1 = \cosh\alpha_I(x_b - x_l) + \sqrt{\kappa_M/\kappa_I}\coth\alpha_M(1 - x_b)\sinh\alpha_I(x_b - x_l)\;,$$

and

$$C_2 = \cosh\alpha_I(x_b - x_l) + \coth\alpha_I x_l \sinh\alpha_I(x_b - x_l)\;.$$

For the case $x_l > x_b$ we get a similar equation by replacing $x$ by $1 - x$, $x_b$ by $1 - x_b$, and $x_l$ by $1 - x_l$, and interchanging $\kappa_I$ and $\kappa_M$. Using this and Eq. (S3) to calculate $\delta T_b$, we plot $V_{te} = -\Delta S_{IM}\delta T_b$ along with the measured emf $V_0$ at both temperatures for Device 2 in Fig. 3d. We take[3] $\kappa_I = 3.5$ W/m/°C, $\kappa_M/\kappa_I = 2$ (the result is weakly sensitive to this ratio, which was estimated at 1.6 in Ref. 3), $L = 18$ μm, $A = 0.2$ μm$^2$, and $\Delta S_{IM}$ = -330 μV/°C (see text), and treat $T_P$ and $\beta$ as fitting parameters. The best fit is obtained for $T_P \approx 2.8$ °C and $\beta \approx 0.03$ W/m/°C. From these we get $\alpha_I = 3.7$ and $\alpha_M = 2.6$ and $\gamma \approx 0.45$. The value of $\beta$ is consistent with thermal conduction through the air ($\kappa_{air} \approx 0.03$ W/m/°C at 80 °C) between the bottom of the nanobeam which is of width ~1 μm and the substrate at a distance ~1 μm beneath it.

## 5. Calculation of the thermoelectric peak emf vs stage temperature

The peak in $V_{te} = -\Delta S_{IM}\delta T_b$ occurs when $x_l = x_b$ and

$$\delta T_b = \delta T_l = T_P\,\frac{1}{\coth\alpha_I x_b + \sqrt{\kappa_M/\kappa_I}\coth\alpha_M(1 - x_b)}\;,$$

so its dependence on the stage temperature $T_0$ is given by

$$V_{te}(T_0) = V_P\,\frac{1}{\coth\alpha_I x_b(T_0) + \sqrt{\kappa_M/\kappa_I}\coth\alpha_M(1 - x_b(T_0))}\;, \qquad (3)$$

where $V_P = -\Delta S_{IM}\gamma P/\sqrt{\kappa_I \beta A}$ . The solid line in Fig. 3c is a plot of this function using the measured values of $x_b(T_0)$ from Fig. 3a and the values of the other parameters from above (the nanobeam has similar dimensions), except that to best fit the measurements on Device 1 we take $\gamma = 0.6$, somewhat larger than for Device 2. (This makes the peak emf 30% larger in Device 1 than Device 2; the difference could also be in part due to a slightly different cross-sections of the two nanobeams.)